# Plasmon-enhanced chiral absorption through electric dipole-electric quadrupole interaction

Hanwei Wang[1,2], Yang Zhao[1,2,3,4,5]

[1] Department of Electrical and Computer Engineering, University of Illinois Urbana-Champaign, Urbana, IL 61801
[2] Nick Holonyak Micro and Nanotechnology Laboratory, University of Illinois Urbana-Champaign, Urbana, IL 61801
[3] Beckman Institute, University of Illinois Urbana-Champaign, Urbana, IL 61801
[4] Department of Bioengineering, University of Illinois Urbana-Champaign, Urbana, IL 61801
[5] Carl R. Woese Institute of Genomic Biology, University of Illinois Urbana-Champaign, Urbana, IL 61801

E-mail: yzhaoui@illinois.edu

**Abstract**

Enantioselective interactions of chiral molecules include distinct absorptions to opposite-handed circularly polarized light, known as chiral absorption. Traditionally, chiral absorption has been primarily attributed to electric dipole and magnetic dipole interaction with molecular chirality. However, this approach falls short for large molecules that support high-order multipolar components, such as electric quadrupole moment. Here, we introduce a theoretical model to study the chiral absorption of large molecules in the presence of plasmonic nanostructures. This model considers both electric dipole-magnetic dipole interaction and electric dipole-electric quadrupole interaction enhanced by a resonant structure. We numerically study such interactions of the chiral molecular solution in the vicinity of an achiral plasmonic nano-resonator. Our results show the distinct spectral information of the chiral medium on- and off-resonance of the resonator.

## 1. Introduction

Chirality is used to describe molecules or structures that are not superimposable with their mirror images. Enantiomers, a pair of chiral molecules with opposite handedness, may present different biological, physical, or chemical properties and, therefore, have been widely studied. Enantioselective light-matter interactions, such as absorption [1-3], fluorescence [4-7], and optical forces [8-13], have been applied to detect and separate enantiomers. Macroscopically, such chiral interactions arise from the differences in the molecule's interactions with left- and right-handed circularly polarized light, which could be measured by circular dichroism (CD) spectroscopy. However, the CD signal is typically weak for low concentrations of molecules, such as DNAs [14] and proteins [15]. To measure trace amounts of chiral molecules, high quality factor dielectric nano-resonators or metamaterials could be used to capture the enantioselective shift of the resonance frequency [16-20]. Furthermore, plasmonic nanostructures have been employed to enhance chiral absorption through an increased local optical chirality density [21-23]. A traditional understanding of such enhancement is through electric dipole-magnetic dipole (ED-MD) interaction; the surface plasmon resonance enhances the local optical chirality, and thus, increases the chiral dipole absorption of ED-MD interaction [1, 24-26]. However, this understanding overlooks the chiral absorption due to higher-order multipolar absorption, especially when plasmon-induced spatial non-homogeneity of the fields is present [27].

Most biomolecules support non-zero quadrupole moments [28], such as proteins with complex secondary structures [29]. When the molecular dipole is aligned with an external electric field, the electric dipole-electric quadrupole (ED-EQ) interaction may contribute to a non-zero quadrupole moment of the medium [30]. In a general gradient field, such an interaction could result in chiral absorption [27]. However, little is known about how such interaction will contribute to the overall chiral absorption with spatially varying gradient fields, such as those induced by resonant nanostructures. The molecular ED-EQ interaction is sensitive to the field gradient created by localized surface plasmon resonance, which could result in additional chiral absorption, while such enhancement mechanisms and spectral features are expected to be distinct from the well-known ED-MD interaction [24].

In this paper, we develop a theoretical model for a chiral medium that supports both ED-MD and ED-EQ interactions. We study the overall chiral absorption induced by resonance enhancement with a gold metal-insulator-metal nanoparticle. We show that the nanoparticle will not only enhance the molecule's chiral absorption through an increased optical chirality density but also an enhanced ED-EQ interaction by the field gradient. As an analogy to the definition of optical chirality, we introduce an extended optical chirality to measure the gradient field's contribution to chiral absorption. This study provides a new paradigm for selectively enhancing the detection of large chiral molecules exhibiting quadrupole moments.

## 2. Theory

The chiral medium supports a quadrupole moment through ED-EQ interaction, analogous to certain proteins functionalized on a surface. The EQ moment density (electric quadrupole moment per unit volume) is given by [31]

$$Q_{i,j}^{\pm} = \sum_k \gamma_{i,j,k} P_k^{\pm}, \quad (1)$$

where $\pm$ represents the opposite parity of light, for example, being left- and right-handed [1], $\gamma_{i,j,k}$ is the ED-EQ coupling coefficient [30]; i, j, k indicates the Cartesian coordinates; $P$ is the electric polarization vector. For simplicity, we assume the chiral medium has an isotropic ED-EQ coefficient, $\gamma$; and Q only has values in the diagonal terms, i.e., i, j, k in Eq. (1) being the same. Explicitly, $Q_{xx}^{\pm} = \gamma P_x^{\pm}$, $Q_{yy}^{\pm} = \gamma P_y^{\pm}$.

The electric polarization vector (electric dipole moment per unit volume) is influenced by the ED-MD interaction [26]:

$$P_k^{\pm} = \varepsilon_0 \chi E_k^{\pm} - \frac{i\kappa}{c} H_k^{\pm}, \quad (2)$$

where $\chi$ is the linear susceptibility, $\kappa$ is the Pasteur parameter, which describes the ED-MD interaction, c is the speed of light.

Assuming the molecules to be non-magnetic, the magnetization (magnetic dipole moment per unit volume) is given only by ED-MD interaction [26]:

$$M_k^{\pm} = \frac{i\kappa}{c\mu_0} E_k^{\pm}, \quad (3)$$

To simulate the chiral medium, we modify the domain equation. The electric displacement field is contributed by both electric polarization and EQ moment density [32]:

$$D_k^{\pm} = \varepsilon_0 E_k^{\pm} + P_k^{\pm} - \frac{1}{2}\frac{\partial}{\partial k}Q_{kk}^{\pm}. \quad (4)$$

Substituting Eqs. (1)-(2) into (4):

$$D_k^{\pm} = \left[\varepsilon_0(1+\chi) - \frac{1}{2}\varepsilon_0\chi\gamma\frac{\partial}{\partial k}\right]E_k^{\pm} - \frac{i\kappa}{c}\left[1 + \gamma\frac{\partial}{\partial k}\right]H_k^{\pm}. \quad (5)$$

The magnetic flux density is given by magnetic field strength and magnetization as:

$$B_k^{\pm} = \mu_0 H_k^{\pm} + \mu_0 M_k^{\pm}. \quad (6)$$

According to Eq. (3):

$$B_k^{\pm} = \mu_0 H_k^{\pm} + \frac{i\kappa}{c}E_k^{\pm}. \quad (7)$$

Maxwell equations using the revised electric displacement field in Eq. (5) and magnetic flux density in Eq. (7) can describe the properties of the chiral medium. Assuming the medium to be source-free (no free current or charge), they are $\nabla \times \mathbf{E} = -i\omega\mathbf{B}$, $\nabla \times \mathbf{H} = i\omega\mathbf{D}$, $\nabla \cdot \mathbf{D} = 0$, and $\nabla \cdot \mathbf{B} = 0$.

The absorption, given by ED, MD, and EQ, is respectively:

$$A_{ed}^{\pm} = -\frac{1}{2}Re\left(i\omega \sum_k E_k^{\pm *} D_k^{\pm}\right), \quad (8)$$

$$A_{md}^{\pm} = -\frac{1}{2}Re\left(i\omega \sum_k B_k^{\pm *} H_k^{\pm}\right), \quad (9)$$

and

$$A_{eq}^{\pm} = -\frac{1}{2}Re\left(i\omega \sum_k \frac{\partial E_k^{\pm *}}{\partial k} Q_{kk}^{\pm}\right). \quad (10)$$

Particularly substituting Eqs. (1)-(2) into Eq. (10), we get:

$$A_{eq}^{\pm} = \frac{\omega\gamma\varepsilon_0}{2}\sum_k Re\left(\chi \frac{\partial E_k^{\pm *}}{\partial k} E_k^{\pm}\right) - \frac{\omega\gamma}{2c}\sum_k Im\left(\kappa \frac{\partial E_k^{\pm *}}{\partial k} B_k^{\pm}\right), \quad (11)$$

The first and the second terms correspond to the achiral and chiral absorption of EQ, respectively. Specifically, the chiral absorption is correlated with the term $\frac{\partial E_k^*}{\partial k} B_k$, which could be raised by a gradient field. For the dipole absorptions due to ED-MD interactions, the absorption is $A_{ed}^{\pm} =$ $\frac{1}{2}\omega\left(\varepsilon_0\chi''Re(\boldsymbol{E}^{\pm *}\cdot\boldsymbol{E}^{\pm})\pm\frac{2}{c}\kappa''Im(\boldsymbol{E}^{\pm *}\cdot\boldsymbol{B}^{\pm})\right)$ [1]. The chiral EQ absorption in Eq. (11) has a format similar to that of the chiral dipolar absorption. Compared with the standard definition of optical chirality [1], $Im(\boldsymbol{E}^*\cdot\boldsymbol{B})$, we define an extended optical chirality, $C_k^e = \frac{\partial E_k^*}{\partial k}B_k$. The chiral EQ absorption is correlated with the extended optical chirality. Particularly, at the resonance frequency of $\kappa$, if following a Lorentzian line shape, $\kappa$ will be purely imaginary. In that case, the chiral EQ absorption is proportional to the real part of the extended optical chirality.

Assuming the gradient of the magnetic field to be zero, we can derive the ED absorption from Eq. (8) as

$$A_{ed}^\pm = \frac{\omega\varepsilon_0\chi''}{2}\sum_k Re(E_k^{\pm *}E_k^\pm) - \frac{\omega}{2c}\sum_k Re(\kappa E_k^{\pm *}H_k^\pm) + \frac{\gamma\varepsilon_0\omega}{4}\sum_k Re\left(\chi E_k^{\pm *}\frac{\partial E_k^\pm}{\partial k}\right), \quad (12)$$

where the first term represents the achiral absorption, the second term represents the chiral absorption due to ED-MD coupling, and the third term represents the chiral absorption due to ED-EQ coupling. We can see that when the gradient field is formed by the plasmon resonance, the chiral absorption due to ED-EQ coupling and the chiral absorption due to ED-MD coupling have opposite signs, resulting in reduced chiral absorption.

The MD absorption can be derived from Eq. (9) as

$$A_{md}^\pm = \frac{\omega\mu_0}{2}\sum_k Re(H_k^{\pm *}H_k^\pm) - \frac{\omega}{2c}\sum_k Re(\kappa E_k^{\pm *}H_k^\pm), \quad (13)$$

where the first term represents the achiral absorption and the second term represents the chiral absorption due to ED-MD coupling, which can be enhanced by the plasmon resonance due to the increased local field strength.

Please note that the chiral absorption given by EQ is induced not only by ED-EQ interaction but also by ED-MD interaction. By comparing Eqs. (11-13), it is seen that the chiral absorption of EQ must require a field gradient while MD and ED do not; as a result, the influence of the plasmon resonance on chiral absorption is most significant for EQ among the three. The chirality of the molecule is still given by the ED-MD interaction, characterized by $\kappa$; however, as the ED is coupled to the EQ, an additional chiral absorption is enhanced by the extended chirality $C^e$. In the following, we show numerical examples to elucidate this phenomenon quantitatively.

## 3. Numerical Model

We simulate the chiral absorption associated with ED, MD, and EQ, which is enhanced by a plasmonic nanostructure. To incorporate the properties of the chiral medium, we modified the system equation in the RF module of COMSOL Multiphysics. As shown in Figure 1(a), the nanostructure consists of a sphere in the center, surrounded by a ring. The radius of the inner sphere, denoted as $r_1$, is 57 nm, and the radius of the surrounding ring, denoted as $r_2$, is 100 nm. The separation between the sphere and the ring is 10 nm in the radial direction. The plasmon resonance is supported at the gap between the two metallic nanostructures, forming a metal-insulator-metal nanoparticle; a similar structure has been shown to enhance enantioselective light-matter interactions [9]. The material of the nanostructure is gold, which has properties following [33]. The simulation domain is spherical with a radius of 1 $\mu m$; it is surrounded by a perfectly matched layer with a thickness of 300 nm. The gold nanostructure is placed in a homogeneous chiral medium. The chiral medium is modeled by considering both ED-MD interaction and ED-EQ interactions, using the modified domain equation. Specifically, the electric displacement field, $\boldsymbol{D}$, is modified according to Eq. (5); and the magnetic flux density, $\boldsymbol{B}$, is modified according to Eq. (7).

The chiral absorption is defined as the differential absorption with the right (+) circularly polarized light and left (-) circularly polarized light:

$$A_c = A^+ - A^-. \quad (14)$$

The total absorption is defined as

$$A_{tot} = (A^+ + A^-)/2. \quad (15)$$

The chiral absorption power density, defined as the chiral absorption per unit volume, which is associated with ED, MD, and EQ as calculated by Eqs. (8-10), is depicted in Figure 1(c). The pronounced chiral absorption power density for the ED and EQ is located at the gap between the sphere and the ring, similar to the electric field hotspot shown in Figure 1(a). Such a phenomenon presents a plasmon enhancement of the chiral absorptions by the nanostructure.

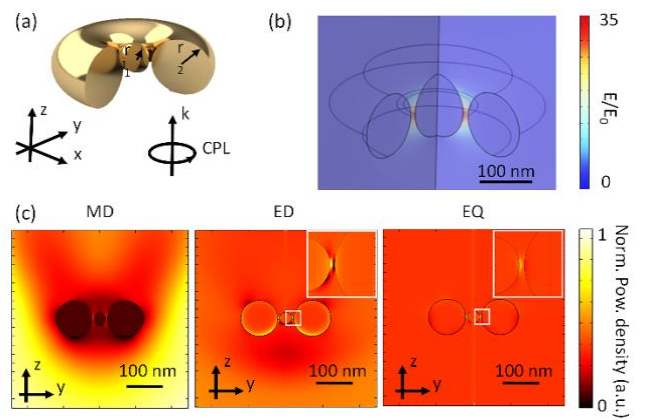

Figure 1. Schematic illustration of plasmon-enhanced chiral absorption: (a) Plasmonic nanostructure with a circularly polarized incidence, (b) Electric field, (c) Chiral absorption power density resulting from magnetic dipole (MD), electric dipole (ED), and electric quadrupole (EQ),

respectively. The simulation uses a unity electric field strength of the incident light (1 V/m). The insets show the Zoom-in plots of the highlighted region for ED and EQ.

We model the Pasteur parameter, $\kappa$, with a Lorentzian line shape with a central wavelength of 770 nm and a full width at half maximum (FWHM) of 50 nm (Figure 2(a)), $L = \frac{FWHM}{FWHM^2+(\lambda-\lambda_0)^2} + \frac{i(\lambda-\lambda_0)}{FWHM^2+(\lambda-\lambda_0)^2}$, where $\lambda$ is the wavelength of the incident light, $\lambda_0$ is the resonance wavelength. Although we considered $\gamma$ to be non-dispersive for simplicity, the method can be extended to dispersive $\gamma$ by replacing $\gamma$ with a wavelength-dependent complex value. The main conclusions illustrated in this paper remain unchanged. The achiral and chiral absorption induced by MD, ED, and EQ are shown in Figure 2(b). It is observed that the EQ plays a significant role in dominating the chiral absorption. It is important to note that in molecules lacking ED-EQ interaction, the chiral absorption of ED and MD is identical. The difference in them, as shown in Figure 2(b), is attributed to the ED-EQ interaction. When the nanostructure is on-resonance, the chiral absorption associated with the ED diminishes to nearly zero, whereas those of the EQ and MD both increase. Notably, the rate of increase is especially significant for the chiral absorption contributed by the EQ.

The effects of the Pasteur parameter, $\kappa$, and the ED-EQ coupling coefficient, $\gamma$, are shown in Figure 2(c). The achiral absorption is affected by $\gamma$ due to quadrupole absorption and is not significantly influenced by $\kappa$. Conversely, the chiral absorption associated with the MD primarily depends on $\kappa$. In contrast, the chiral EQ absorption is sensitive to both ED-MD and ED-EQ interactions, aligning with the theoretical predictions presented in Eq. (11). Furthermore, the pronounced ED-EQ transition markedly impacts the chiral absorption of the ED, as evidenced by a decreased value when $\gamma$ approaches to 100 nm.

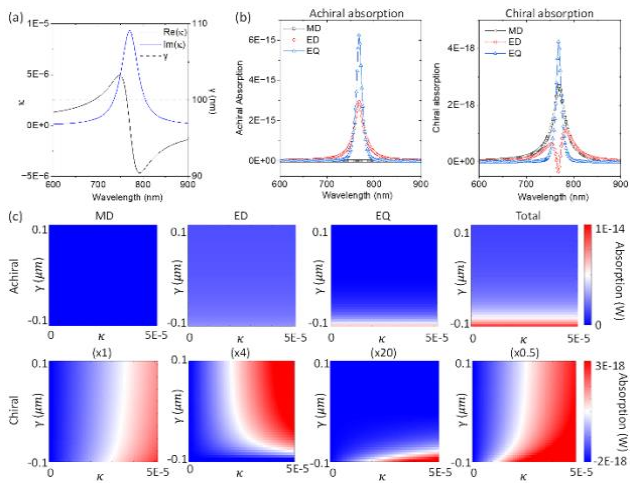

Figure 2. Influence of multipolar interaction intensity: (a) Spectrum of the real and imaginary parts of the electric dipole (ED)-magnetic dipole (MD) coupling coefficient $\kappa$, and a non-dispersive electric dipole (ED)-electric quadrupole (EQ) coupling coefficient, $\gamma$. (b) Spectrum of the achiral and chiral absorption attributed to the ED, MD, and EQ, respectively. In the simulation, $\kappa$ has a peak value of $5E-5$ and a FWHM of 50 nm, and the ED-EQ coupling coefficient $\gamma$ ranges from -0.1 $\mu m$ to 0.1 $\mu m$. (c) Maps showing the variation of each component's chiral and achiral absorption with respect to $\kappa$ and $\gamma$ with on resonance condition (inner sphere radius, $r_1$= 57 nm).

We further vary the geometry of the nanostructure to manipulate the plasmon resonance. At a wavelength of 770 nm, the resonance is achieved with an inner sphere radius of 57 nm. As shown in Figure 3(a), approaching the resonance condition increases the chiral absorption attributed to the EQ and decreases that of the ED. This transition indicates that the plasmon resonance not only amplifies the total chiral absorption but also alters its composition, which may correspond to the unique properties of the chiral medium. Specifically, the plasmon resonance significantly boosts the chiral absorption of the EQ. As shown in Figure 3(b), the strong chiral EQ absorption is correlated with the extended optical chirality induced by the plasmon resonance, corroborating Eqs. (11-13) of our theory.

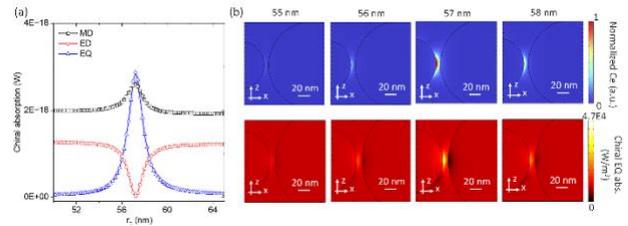

Figure 3. Influence of the nanostructure's geometry on chiral quadrupole (EQ) absorption. (a)Variation of chiral EQ absorption with the inner sphere radius, $r_1$. (b) Relationship between extended optical chirality, $C^e$, and chiral EQ absorption power density for different values of $r_1$. In the calculations, $\kappa$ has a peak value of $5E-5$ and a FWHM of 50 nm, $\gamma = 0.1\ \mu m$, and the radius of the outer gold ring $r_2$ = 100 nm.

The on- and off- resonant states of the nanostructure may lead to variations in the origins of the chiral absorption, thereby affecting the spectral information of the chiral medium. As shown in Figure 4(a), in the off-resonant condition, the chiral absorption attributable to the ED and MD is significantly stronger than that of the EQ. The chiral absorption of the ED and MD is primarily influenced by $\kappa$, and the effect of $\gamma$ on the chiral absorption of the EQ is symmetric in both positive and negative directions. However, when the nanostructure is on-resonance, the chiral absorption associated with the EQ becomes predominant (Figure 2c). The sign of $\gamma$ also exerts a significant impact on the total chiral absorption. The asymmetry between positive and negative $\gamma$ values indicates a strong interplay between ED-EQ and ED-MD interactions. This effect is also observable in the dissymmetry factor, defined as the ratio between the chiral absorption and the total absorption: $g = 2\frac{A_{RCP}-A_{LCP}}{A_{RCP}+A_{LCP}}$, where $A_{RCP}$ are $A_{LCP}$ are absorptions under right- and left-handed circularly polarized light, respectively. As shown in Figure 4(b), the dissymmetry factor g primarily correlates with $\kappa$ in the off-resonant state, indicating that the spectral information related to the EQ is not observable. Conversely, when the nanostructure is on-resonance, the dissymmetry factor g becomes related to both $\gamma$ and $\kappa$. The extended chirality $C^e$ induced by the nanostructure introduces a new mechanism of chiral absorption through the ED-EQ coupling, leading to not only an enhancement but also additional spectral information of the chiral medium.

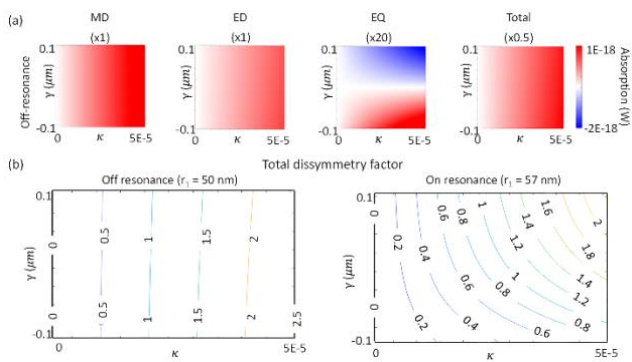

Figure 4. Comparison between the off-resonant and resonant nanostructures. (a) Chiral absorptions attributed to the magnetic dipole (MD), electric dipole (ED), and electric quadrupole (EQ) for the off-resonant nanostructure ($r_1 = 50$ nm). (b) Contour maps depicting the total dissymmetry factor g for both off-resonant and resonant conditions of the nanostructures.

## 4. Conclusion

This study develops a method for simulating a homogenous chiral medium that supports ED, MD, and EQ moments. Through theoretical analysis and numerical simulations, we investigate the enhancement effect of a plasmonic nanostructure on the chiral medium. Our findings reveal that nanostructures, such as a gold nanoparticle, interact with chiral media not merely by amplifying the optical chirality but also by enhancing quadrupole chiral absorption through field gradients. When the nanostructure is on-resonance, the chiral absorption becomes sensitive to both the coupling of ED-MD and ED-EQ. The overall chiral absorption or dissymmetry factor g encapsulates information pertaining to both polarizabilities. This method extends our understanding of the role of ED -EQ interaction in chiral absorption and paves the way for analyzing large chiral molecules such as proteins or DNA nanostructures in plasmon-enhanced CD spectroscopy.

## Acknowledgments

This work was supported in part by the Dynamic Research Enterprise for Multidisciplinary Engineering Sciences (DREMES) at Zhejiang University and the University of Illinois at Urbana-Champaign and by the Jump ARCHES endowment through the Health Care Engineering Systems Center.